\documentclass[12pt,oneside,reqno]{amsart}

\numberwithin{equation}{section}

\usepackage{amssymb}
\usepackage{graphicx}
\usepackage{bm}
\usepackage{footnote}

\begin{document}

\title{On origin and statistical characteristics \\ of 1/f-noise}

\author{Yuriy E. Kuzovlev\,,\, German N. Bochkov}
\address{Donetsk Institute for Physics and Technology of NASU,
ul.\,R.\,Luxemburg 72, Donetsk 83114, Ukraine\,\,;\,\, %
\newline %
\,\,\,\, Nijnii Novgorod State University, pr.\, Gagarin 23, 603950 %
Nijnii Novgorod, Russia}
\email{kuzovlev@fti.dn.ua \,;\,\, histat@rf.unn.ru}





\begin{abstract}
We suggest some principal ideas on origin, statistical properties %
and theoretical description of 1/f-noise  %
exactly as they for the first time were expounded in our preprint %
published in Russian in 1982, and supplement them with  %
short today's comments and selected references, with %
wish to support improvements of present generally poor %
ideologic and mathematical base of the 1/f-noise theory. %
\end{abstract}


\maketitle

\baselineskip 18 pt

\markboth{}{}


{\it ``Many things are incomprehensible not because %
our notions are poor

but because these things are not in a frame of our notions''

(Koz'ma Prutkov) }

\,\,\,

    PREFACE

\,\,\,

As far as we know, the present state of the %
1/f-noise problem has no principal differences %
from that thirty years ago when our preprint had %
appeared (Preprint NIRFI No.157, published in 1982 by the %
Scientific-Research Radio-Physics Institute in Nijnii Novgorod %
in Russian Federation). Of course, this is said not about %
varieties of objects under experimental investigation %
but about conceptual level of their theoretical %
interpretation. Reading of today's scientific %
literature shows us that the interpreters %
as before are entangled in prejudices %
cultivated by more than centennial history %
of hypotheses, assumptions and approximations %
in statistical physics and kinetics. %
We in our preprint just made first attempt to %
disentangle. Therefore we think that it still may %
be useful for interested readers, all the more that %
it affected all our later works (first of all \cite{pjtp,bk12,bk3,pr2}).

Later, we have discovered N.\,Krylov's book \cite{kr} %
which is devoted to disclosure of the prejudices %
and gave us principal justification of our own findings. In turn, our own %
investigations of ``molecular Brownian motion'' %
(\cite{i1} - \cite{p1209}, \cite{p1207}, \cite{bkn}) %
and ``electron  Brownian motion'' %
(\cite{i2}, \cite{kmg} - \cite{p1202}, %
\cite{p1207}, \cite{bkn}) %
in statistical mechanics gave confirmation of %
Krylov's ideas. As the result, in particular, %
now we have at our disposal more correct and %
formally substantiated terminology than thirty %
years ago. Nevertheless, our translation of the %
preprint is as much ``one-to-one'' as possible. %
Some necessary corrections and additions can be %
found in the mentioned %
references and in comments below designed also %
as references but with mark \checkmark\,.

\,\,\,

\tableofcontents

\section*{Abstract}

It is shown that thermodynamically equilibrium Brownian motion %
of charge carriers always possesses low-frequency fluctuations %
in diffusivities of the carriers and thus in power spectrum of ``white'' %
noise of the conducting media, with correlation function of the %
fluctuations decaying by logarithmic law $\,(\ln\, t/\tau_0)^{-1}\,$ %
while their spectrum is of $\,1/\omega $\, type. %
Similar fluctuations are peculiar to electric conductivity and %
current in non-equilibrium state (under external field). %
It is shown that these fluctuations are not related to some slow %
processes (and macroscopic relaxation times). %
Statistical characteristics of the 1/f-noise are completely %
determined by microscopic parameters of ``fast'' random %
motion of the carriers. The 1/f-type spectrum reflects %
absence of long-living correlations at this random motion.

An exhaustive information about spectrum and magnitude %
of the 1/f-noise is contained in fourth-order cumulant %
function of equilibrium current fluctuations.

The suggested theory estimates the 1/f-noise %
intensity in agreement with experimental data %
and explains origin of the empiric ``Hooge constant''.

\,\,\,

\section{Introduction}

\subsection{Experimental data and empiric relations}


Electric charge transport in various media and systems is %
accompanied by characteristic low-frequency noise, - %
so called excess noise, or flicker noise, - %
which is known almost as long as usual white noise, %
but still has no theoretical explanation. %

Surprising peculiarity of flicker noise is that %
its spectral power density is fast increasing %
as frequency decreases, approximately  %
by law $\,\omega^{-\alpha}$\,, and does not %
show tendency to saturation down to %
minimal measurable frequencies\, %
$\sim 10^{-6}\div 10^{-7}\,$ Hz\,. In most %
cases the exponent $\,\alpha\,$ is close to unit,\, %
$\alpha\approx 1$\,, and then it is said about 1/f-noise. %

Intensity of flicker fluctuations of electric current %
$\,J(t)\,$ or voltage in weakly non-equilibrium states %
(in Ohmic regime) is proportional to squared mean current, %
therefore flicker noise is usually considered as result of %
fluctuations of resistance or conductance. %
Main relationships of electric current  1/f-noise in homogeneous %
conducting media are reflected by approximate empiric %
``Hooge formula'' (for details see reviews [1-2]): %
\begin{eqnarray}
S_J(\omega)\,=\, \overline{J}^2\,\, \frac %
{2\pi \,a}{N\,\omega}\,\, \,\, \label{fh}
\end{eqnarray}
Here $\,S_J(\omega)\,$ is spectral power density of %
current fluctuations,\, $\, \overline{J}\,$ is mean current,\, %
$N\,$ is number of charge carriers in a sample of media, %
and $\,a\,$ is dimensionless quantity.

Formula (\ref{fh}) reflects one more surprising property %
of 1/f-noise:\, indifference of shape of its spectrum %
to system's geometry and sizes, the so-called %
``zero-dimensionality'' of 1/f-noise. %

Formula (\ref{fh}) usually gives satisfactory description %
of 1/f-noise in semiconductors, solid and liquid metals, %
electrolytes [1]. In case of semiconductors %
the quantity $\,a\,$ is almost independent on %
temperature $\,T\,$ and the number $\,N$\,. %
The latter observation is evidence of statistical %
independence of contributions from  particular %
charge carriers into full 1/f-noise. %
Interestingly, in various intrinsic %
(weakly doped) semiconductors $\,a\,$ has nearly %
same order of magnitude, $\,a\sim 0.001\,$ %
(for the first time this fact was noticed by Hooge [3]). %
At high enough temperatures, values\, %
$a\sim 0.001\, \div \, 0.01\,$ characterize also %
metals, though there $\,a\,$ is temperature-dependent [1]. %
For electrolytes also one can find  $\,a\sim 0.001$\, %
\checkmark \cite{c1}. 

There are three sorts of significant %
differences of 1/f-noise level from that %
responding to $\,a\sim 10^{-3}$\,. %
In strongly doped semiconductors the noise %
is much weaker. There, according to Vandamme and Hooge,\, %
$a\approx 10^{-3}\,(\mu/\mu_0)^2$\,, where $\,\mu\,$ is mobility %
of carriers and $\,\mu_0\,$ their mobility in pure material. %
In inhomogeneous systems (among which, seemingly, %
one should rank also very thin films and wires) %
the noise may be much greater. In metals at %
comparatively  low temperatures one can observe %
``temperature 1/f-noise'' [1,2,4,5]. It is characterized %
by $\,a$\,s dependence on conductivity's temperature %
coefficient and sample's thermal contact with %
surroundings. Perhaps, as was noticed in [5], %
in metals the noise represents superposition of %
such ``temperature 1/f-noise'' and the above %
mentioned noise which dominates at higher %
$T\,$ ($\,T\gtrsim 150^\circ\,$K\,).

In general, the picture of flicker noise in metals %
looks much more complicated than in semiconductors. %
In many cases measurements of the exponent %
$\alpha\,$ show values visually different from %
one (\,$0.8\lesssim\alpha\lesssim 1.2\,$), %
so that formula (\ref{fh}) appears too rough. %

\subsection{Correlation experiments. %
``Zero-dimensionality'' of 1/f noise }

 ``Zero -dimensionality'' of 1/f-noise %
expressively manifests itself in so-called %
correlation experiments:\, %
small neighbor regions of same conducting sample %
produce uncorrelated contributions to its 1/f-noise. %
This seems striking, since we speak about %
extremely slow, in microscopic time scale,  fluctuations. %
Indeed, 1/f-type spectrum can be expanded %
into sum of Lorentzians,
\begin{eqnarray}
\frac 1\omega\, \rightarrow\, %
\frac2\pi \int_{\Omega_0}^\infty %
\frac {\Omega}{\omega^2 +\Omega^2}\,=\, %
\frac 1\omega \,\left( \frac 2\pi\, %
\arctan\,\frac \omega{\Omega_0} \right)\,\,, \, \nonumber
\end{eqnarray}
were, as was mentioned, %
$2\pi/\Omega_0 > 10^6\,$s\,. And we may assume that %
there is some real fluctuation mode beyond either %
Lorentzian, with relaxation time $\,2\pi/\Omega\,$.
However, it is hard to imagine flutuations' mechanism %
what would be space-local but at that lead to %
time scales up to $\,10^6\,$s (if not greater) \checkmark \cite{c2}.

Though, the ``temperature 1/f-noise'' in metals %
possesses spatial correlations [4]. %
Therefore Voss and Clarke concluded in [4] that it %
is caused by mere thermodynamic fluctuations of temperature %
``modulating'' conductivity. Now, one can state %
that this conclusion was wrong. %
The theory of temperature fluctuations, %
consistent with principles of statistical thermodynamics, %
does not lead to 1/f-type spectra [1,2,4]. %
At the same time, there are no doubts that the %
``temperature 1/f-noise'' is closely related to thermal processes. %

It is easy to accept the complexity of situation with metals. %
In metals, electrons are not only charge carriers but also %
main heat carriers, so that electric and thermal processes %
are mutually entangled.

\subsection{Thermodynamically equilibrium 1/f-noise %
and fourth-order cumulant of current fluctuations}

The accumulated experimental data give evidence that %
1/f-noise is by its nature thermodynamically equilibrium [1,2]. %
This principally important circumstance means that %
unknown processes, what are responsible for 1/f-noise, %
take place first of all in equilibrium state, %
although do not manifest themselves in correlation function %
and spectrum of electric current or e.m.f.

Nevertheless, in some things these processes must be %
reflected even in equilibrium case. It is not hard %
to guess that they must lead to flicker fluctuations %
of intensity of equilibrium white noise, $\,S(t)$\,. %
This is prompted already by the Nyquist formula:\, %
$S=2Tg\,$,\, if one interprets 1/f-noise as consequence %
of fluctuations of conductance, $\,g(t)$\,. Voss and Clarke %
measured fluctuations of white noise power in thin metal film %
and really found that they have 1/f-type spectrum [4]. %
This phenomenon also can be termed 1/f-noise (equilibrium %
now just in literal sense). In out-of-equilibrium system %
under external field it transforms into current (and voltage) %
fluctuations which will be termed below current 1/f-noise.

It should be underlined that the fluctuating power %
$S(t)$\,, in contrast to $\,J(t)\,$, is especially %
phenomenological characteristics of noise. %
While $\,J(t)\,$ always can be written as a function of %
microscopic dynamical variables of a system, %
$S(t)\,$ can not be expressed by dynamical language %
neither through these variables nor through fluctuations %
of thermodynamical quantities:\, %
temperature, chemical potential, etc.,\, - %
characterizing system's quasi-equilibrium states. %
The matter is that the power (power spectral density) %
of white noise represents kinetic quantity. %
Its definition should involve statistical averaging over %
ensemble and, besides, integration (averaging) over time. %
But in such way one can rigorously introduce only mean %
value\,\footnote{\, Subscript ``0'' at angle brackets %
indicates that averaging is made over equilibrium ensemble.}
\[
\langle \,S(t)\,\rangle_0\,\equiv\, S_0\,= %
\int_{-\infty}^\infty \langle \,J(t)\,J(0)\,\rangle_0\,\, dt\,\,, \,
\]
while fluctuations of $\,S(t)\,$ have no strictly definite %
dynamical sense. The said concerns also fluctuations of %
conductance $\,g(t)\,$ and other kinetic quantities.

Then, how we can rigorously describe fluctuations %
of the power and conductance? It is very simple issue. %
Since  $\,\langle \,S(t)\,\rangle_0\,$ is connected to %
quadratic forms, in respect to current or voltage, %
fluctuations of $\,S(t)\,$ are corresponded by current's statistical %
moments of fourth order (and higher orders). %

Hence, it is necessary to consider fourth moment %
of current's fluctuations:
\begin{eqnarray}
\langle \,J(t_1)\,J(t_2)\,J(t_3)\,J(t_4)\,\rangle_0\,=\, %
\langle \,J(t_1)\,J(t_2)\,\rangle_0\, %
\langle \,J(t_3)\,J(t_4)\,\rangle_0\, +\, %
\, \nonumber \\ +\, %
\langle \,J(t_1)\,J(t_3)\,\rangle_0\, %
\langle \,J(t_2)\,J(t_4)\,\rangle_0\, +\, %
\langle \,J(t_1)\,J(t_4)\,\rangle_0\, %
\langle \,J(t_2)\,J(t_3)\,\rangle_0\, +\, %
\nonumber \\ \,+\, %
\langle \,J(t_1)\,,\,J(t_2)\,,\,J(t_3)\,,\, %
J(t_4)\,\rangle_0\,\, \, \label{f1_2} %
\end{eqnarray}
For it always there is a rigorous dynamical expression. %
The brackets with commas inside on the right in (\ref{f1_2}) %
mean fourth-order cumulant of current. %
Equality in (\ref{f1_2}) is known general cumulant expansion %
of fourth moment at $\,\langle J(t)\rangle_0 =0\,$. %

We have to underline that information about the power fluctuations %
and equilibrium 1/f-noise is hidden in the last term of (\ref{f1_2}), %
i.e. fourth cumulant, which characterizes non-Gaussianity of current %
fluctuations. The equilibrium white noise correlation function %
$\langle J(t)\,J(t_2)\rangle_0 \,\,$ is microscopically fast decaying %
under increase of $\,|t_1-t_2\,$. Therefore low-frequency processes %
are reflected by only last term of (\ref{f1_2}). %

From here an important consequence does follow, %
that in statistical description of 1/f noise it is %
very necessarily to take into account non-Gaussianity %
of current fluctuations (even if it is very weak in one or %
another case)\,\footnote{\, %
Notice that, for instance, Gaussian noise with randomly %
varying intensity becomes non-Gaussian random process. %
For more about analysis of noise non-Gaussianity see %
also:\, M.Nelkin and A.M.S.Tremblay,\, J. Stat. Phys.\, %
{\bf 25} 253 (1981).}. %
Of course, not current only, but also any other physical random %
process always is more or less non-Gaussian. %
Here we meet the situation when this statement is of %
principal importance.

\subsection{Fluctuations of mobility of carriers}

 A number of authors, basing on analysis of experiments, %
have come to conclusion that primary source of 1/f-noise %
is fluctuations of mobilities of carriers [1,2,6]. %
The Hooge formula appears if one assumes that particular mobilities %
fluctuate independently one on another with spectrum
\begin{eqnarray}
S_\mu(\omega)\,=\, \frac %
{2\pi \,a}{|\omega |}\, \mu^2 \,\,, \,\, \label{f1_3}
\end{eqnarray}
where $\,\mu\,$ is mean value of mobility,\, $\,a\approx 10^{-3}\,$. %
This model allows also empirical description of 1/f-noise %
in non-uniform structures:\, in various contacts, %
$p-n-$junctions and other semiconductor devices %
(and, seemingly, in electron emission), moreover, %
even in non-Ohmic regime [1]. %
There, as one can think, 1/f-noise arises because of
fluctuations of carriers' flow to a structural change in turn %
caused by mobility fluctuations.

In applications to semiconductors, the hypothesis of %
mobility fluctuations in many cases meets %
competitive  hypothesis about fluctuations in number %
of carriers due to slow tunnel transitions %
to near-surface states and back [1,2,7]. In this model it is %
easy to obtain 1/f-type spectrum as the sum of Lorentzians %
(thogh saturating at very low frequencies), but it is hard to %
obtain numeric estimates. This model contradicts to %
observed bulk character of 1/f-noise. Besides, special %
experiments with hot carriers also give evidences %
in favor of hypothesis of mobility fluctuations [6].

Then, natural question about physical origin of the %
mobility fluctuations does arise. %
If ascribing their ground to some slow processes %
in crystal lattice (e.g. fluctuations %
in number of phonons \checkmark \cite{c3}, 
as proposed in [1]), one would have to expect mobility %
fluctuations of different carriers to be correlated one %
with another. But experiments say about the opposite. %
If, however, we state absence of inter-carrier correlations, %
then the known ``zero-dimensionality'' paradox arises:\, %
why it is possible to observe flicker correlations %
much longer than mean time of residence of charge carriers %
in a small sample? %
(It seems as if carrier leaves a sample but nevertheless %
its correlations, associated with 1/f noise, stay there.) %
Up to now in the literature none mechanism of the mobility %
fluctuations was suggested. Evidently, the %
mentioned paradox strongly complicates %
the problem \checkmark \cite{c4}. 

\subsection{Principal statements of present work. %
1/f-noise as result of absence of long-living correlations}

In spite of many-year experimental investigations %
and many attempts of theoretical explanation, %
1/f-noise still remains mysterious [1,2]. %
It is all the more strange in view of that %
in other respects the systems under question do not %
display mysterious effects what could be %
associated  with 1/f-noise.

Various investigators' efforts are invariably %
directed either to search of some new ``slow'' %
physical mechanisms, which would  lead to wide set %
of very large relaxation times (or correlation times, %
``life-times'', etc.) and flicker spectrum of %
conductance fluctuations, or to ``approbation'' in %
this sense of already known mechanisms, such as, %
for example, fluctuations of temperature and %
carriers concentration (see e.g. [2,4,8-11]). %
This way does not lead to success.

In the present work, exactly opposite approach to the %
1/f-noise problem is suggested and substantiated, %
allowing to explain this physical phenomenon. %
It is shown that 1/f-noise can be connected just to %
absence of macroscopically large relaxation times %
and caused not by specifically slow processes %
but by Brownian motion of charge carriers in itself, %
i.e. the same ``fast'' microscopic processes only %
which produce diffusion and white noise.

We show that Brownian motion (diffusion) of carriers %
always is accompanied by flicker fluctuations of %
diffusivity and mobility (white noise power and conductance), %
which naturally and inevitably arise in the %
course of diffusion irrespective to its concrete microscopic %
mechanism, without any slow perturbations (as e.g. %
random changes of thermodynamical conditions of diffusion). %
The theory to be expounded below not only yields 1/f-type %
spectrum  but also ensures correct estimate of %
level of 1/f-noise and explains magnitude of the empirical %
``Hooge constant'', $\,a\sim 10^{-3}\,$.

The key, principal, idea of our approach is that 1/f-type %
spectrum is not consequence of real long-living correlations, %
but, oppositely, results from absence of such %
correlations, indifference of system to random deviations %
of ``rate'' of carrier's diffusion  from its mean regime. %
Such deviation is not suppressed by %
backmoving forces, since the only its consequence %
is mere spatial displacement of carrier, that is %
system's transition into a state which is  identical %
in thermodynamical sense to initial state. %

Just in such way, likely, one can explain the %
observed (see above) mobility fluctuations. %
The formulated idea brings also solution of %
the ``zero-dimensionality paradox''. %
Since in reality carrier motion is  %
free of any long-living correlations %
(has no long memory about the past), there is no need %
to speak about %
breaking of such correlations\,\footnote{\, %
It is clear that, very generally, %
indifference of a system to some spontaneous %
deviations and their accumulation with time %
(Brownian displacement of a carrier, in our case) %
can be formally treated both as absence of correlations %
and presence of infinitely long correlations, %
although physically the first of these statements %
is true. Formal (but not essential!) analogy is given %
by a superconductor with non-decaying currents. %
} %
under replacements of one carriers in small %
``noising'' sample by others (by similar reasons, %
the carriers' generation-recombination processes %
do not destroy the 1/f-noise) \checkmark \cite{c5}. 

\section{Brownian motion and equilibrium 1/f-noise. %
Fluctuations of white noise power}

\subsection{Phenomenological description of current fluctuations}

Let us consider a simple physical situation as follows.  %
Taking a sample of conducting medium, let us %
short-circuit it by closing it into ring. %
We are interested in thermodynamically equilibrium %
fluctuations of electric current $\,J(t)\,$ in this closed circuit. %
We shall assume that the medium is statistically %
homogeneous and that charge carriers move statistically %
independently one on another (which usually agrees with %
real situation in semiconductors). Then it is sufficient to %
consider random walk of one separate carrier along the %
ring-like circuit.

Introduce designation $\,v(t)\,$ for random velocity of carrier %
in the ring's direction (thus $\,v(t)\,$ will be scalar), and %
\begin{eqnarray}
r(t)\,=\, \int_0^t v(t^\prime)\, dt^\prime \,\, \,\, \label{f2_1}
\end{eqnarray}
for its path during time $\,t\,$. %
Notice that according to (\ref{f2_1}) $\,r(t)\,$ counts %
(with plus or minus sign) any complete rotation %
around the circuit. In the same sense we shall treat %
displacement $\,r(t)\,$ under charge transfer by not free %
but localized carriers (hopping conductivity) %
when $\,v(t)\,$ is less natural characteristics of the motion. %
By this $\,r(t)\,$'s definition, conveniently, %
in a stationary state $\,r(t)\,$ is random %
process with uniform increments (since $\,v(t)\,$ is %
stationary process). In other words, $\,r(t)\,$ represents %
unbounded diffusion, although the system under consideration %
is essentially bounded. In the very beginning we would like underline %
that our results will be quite insensitive to sizes %
of the system (length of the ring).

Of course, $\,v(t)\,$ always is more or less non-Gaussian random process. %
The simplest usually exploited model of diffusion %
ignores this obvious physical fact and assumes %
$v(t)\,$ and $\,r(t)\,$ Gaussian. This can be justified %
when considering carriers concentration fluctuations %
(just they are taken in mind in literature %
when speaking about ``diffusive noise'', %
``diffusive mechanisms'', etc.) but not in analysis %
of electric noise caused by random wandering of carriers %
(in our system total number of ``noising'' carriers %
does not change during diffusion). %
As already was pointed out, in Gaussian model %
intensity of noise is a priori constant.

Full statistical information about fluctuations of %
$v(t)$, $\,r(t)$, $\,J(t)\,$ in stationary equilibrium  state %
is contained in characteristic functions (CF) %
\begin{eqnarray}
\Theta_t(ik)\,\equiv\, \langle\,%
e^{\,ik\,r(t)}\, \rangle_0 \,\,, \,\label{f2_2}\\ %
\Theta_t(iu)\,\equiv\, \langle\,%
e^{\,iu\,Q(t)}\, \rangle_0 \,\,, \,\,\,\,\, %
Q(t)\,\equiv\, \int_0^t J(t^\prime)\, dt^\prime \, %
\,\, \label{f2_3}
\end{eqnarray}
Here $\,k\,,\,u\,$ are arbitrary probe parameters. %
It should be underlined that in principle one always %
can concretize rigorous microscopic expressions for %
these mathematical objects, therefore use of CF by itself %
does not presume any approximations. %
However, in practice one can not do without %
some assumptions leading to simple enough %
stochastic model.

In Gaussian model, as is well known,
\begin{eqnarray}
\Theta_t(ik)\,=\, e^{-\,D\,k^2t}\, \,, \, \,\,\,\,\, %
\Theta_t(iu)\,=\, e^{-\frac 12\, S\,u^2t}\, \,, \, \label{f2_4}
\end{eqnarray}
where $\,D\,$ is diffusivity (diffusion coefficient) and %
$S\,$ is spectral power density of equilibrium current %
noise at zero frequency. Here it is assumed that %
$t\gg \tau_\mu$\,, where $\,\tau_\mu\,$ is correlation %
time of equilibrium noise. More accurate form of (\ref{f2_4}) %
can be presented with the help of functions %
\begin{eqnarray}
 \Delta_t(ik)\,\equiv\, \frac 1t\, \ln\, \langle\,%
e^{\,ik\,r(t)}\, \rangle_0 \,=\, %
\frac 1t\, \ln\, \Theta_t(ik)\,\,, \,\label{f2_5}\\
\Delta_\infty(ik)\,\equiv\, %
\lim_{t\rightarrow\infty} \, \frac 1t\, \ln\, \langle\,%
e^{\,ik\,r(t)}\, \rangle_0 \,\,,\,\, \nonumber
\end{eqnarray}
and similarly for current. %
Then in Gaussian model
\begin{eqnarray}
 \Delta_\infty(ik)\,=\,-D\,k^2\,\,\, \,\label{f2_6} %
\end{eqnarray}
Probability density distribution of the path $\,r(t)$\,, %
\begin{eqnarray}
W_t(r)\,=\,\frac 1{2\pi} \int_{-\infty}^{\infty} %
e^{-ikr}\, \Theta_t(ik)\,dk\,\,, \,\, \label{f2_7} %
\end{eqnarray}
in this model at $\,t\gg \tau_\mu\,$ is
\begin{eqnarray}
W_t(r)\,=\,(4\pi Dt)^{-1/2}\,\exp{\left( %
-\frac {r^2}{4Dt}\right)}\,\, \, \label{f2_8}
\end{eqnarray}

Now let us consider on-Gaussian model which %
takes into account fluctuations of intensity %
of current noise \checkmark \cite{c6}. 
The concept of such fluctuations %
is meaningful only when they are very slow %
from viewpoint of the scale $\,\tau_\mu\,$. %
Therefore their phenomenological description requires %
to use a model treating $\,v(t)\,$ like Gaussian white noise %
with random ``modulation' of intensity $\,2D(t)\,$ %
(after whuch $\,v(t)\,$ becomes non-Gaussian). %
In such model by definition
\begin{eqnarray}
\left\langle\,\exp{\left\{ \int_0^t ik(t^\prime)\, %
v(t^\prime)\, dt^\prime \right\}}\, %
\right\rangle_0\,=\, \nonumber\\ \,=\, %
\left\langle\,\exp{\left\{ - \int_0^t %
D(t^\prime)\, k^2(t^\prime)\, %
dt^\prime \right\}}\, \right\rangle_0^\prime\, %
=\, \, \label{f2_9}\\ =\, %
\exp{\left\{ - D\int_0^t k^2(t^\prime)\, %
dt^\prime \,+\, \frac 12 \int_0^t \!\int_0^t %
K_D(t^\prime - t^{\prime\prime})\, %
k^2(t^\prime )\, k^2(t^{\prime\prime})\, %
dt^\prime dt^{\prime\prime}\,+\, %
\dots\,  \right\} }\,\,, \,\nonumber
\end{eqnarray}
where the first equality corresponds to %
averaging over white noise at fixed function %
$D(t)$\,, the brackets $\,\langle\,\dots\,\rangle_0^\prime $\,
denote averaging over $\,D(t)$\,'s fluctuations,\, %
$D = \langle D(t)\rangle_0^\prime \,$, %
\begin{eqnarray}
K_D(t^\prime - t^{\prime\prime})\,=\, %
\left\langle\, D(t^\prime)\,D(t^{\prime\prime}))\, %
\right\rangle_0^\prime \,-\,D^2\,\equiv\, %
\left\langle\, D(t^\prime)\,,\, %
D(t^{\prime\prime}))\, %
\right\rangle_0^\prime \, \nonumber
\end{eqnarray}
is correlation function of fluctuations of diffusivity %
$\,D(t)$\,,\, the dots replace contributions to CF %
from higher-order cumulants of fluctuations $\,D(t)$\,,\, and %
$k(t)\,$ is arbitrary probe function.

On the other hand, there is general exact expansion of %
logarithm of the CF (\ref{f2_9}) into series over %
cumulants of the velocity\,\footnote{\, %
Angle bracket with comma-separated factors inside %
(``Malakhov's cumulant bracket'') means joint cumulant %
of these factors.}:
\begin{eqnarray}
\left\langle\,\exp{\left\{ \int_0^t ik(t^\prime)\, %
v(t^\prime)\, dt^\prime \right\} }\, %
\right\rangle_0\,=\, \label{f2_10}\\ \,=\, %
\exp{\left\{ \sum_{n=1}^\infty \, %
\frac {i^n}{n!} \int_0^t \langle \,v(t_1)\,,\, %
\dots\,,\, v(t_n)\, \rangle_0\,\, %
k(t_1) \dots k(t_n)\, dt_1 \dots dt_n\, \right\}}\, \,\nonumber
\end{eqnarray}
In fact, this is definition of the cumulants [13,14]. %
In equilibrium %
$\langle v(t)\rangle_0 =0\,$. %
Equating (\ref{f2_9}) to (\ref{f2_10}), in view of %
$\,k(t)\,$'s arbitraryness, one can obtain for %
fourth-order velocity cumulant
\begin{eqnarray}
\langle\, \,v(t_1)\,,\, %
v(t_2)\,,\, v(t_3)\,,\, v(t_4)\, %
\rangle_0\,=\, \label{f2_11} 
4\,\delta(t_1-t_2)\, \delta(t_3-t_4)\,%
K_D(t_1-t_3)\,+\, \\ +\, %
4\,\delta(t_1-t_3)\, \delta(t_2-t_4)\,%
K_D(t_1-t_4)\,+\, %
4\,\delta(t_1-t_4)\, \delta(t_2-t_3)\,%
K_D(t_1-t_2)\, \nonumber
\end{eqnarray}
From here we find (with taking into account that %
$\,\int_0^\infty \delta(t)\,dt\,=\,1/2\,$)
\begin{eqnarray}
\int_0^\infty \!\int_0^\infty %
\langle\, \,v(t)\,,\, %
v(t^\prime)\,,\, v(t^{\prime\prime})\,,\, v(0)\, %
\rangle_0\, \,dt^\prime\, dt^{\prime\prime}\,=\, %
2K_D(t)\,\, \label{f2_12}
\end{eqnarray}
Clearly, in analogous model of fluctuations %
of current $\,J(t)\,$ for its fourth-order cumulant %
appearing in (\ref{f1_2}) we must obtain
\begin{eqnarray}
\int_0^\infty \!\int_0^\infty %
\langle\, \,J(t)\,,\, %
J(t^\prime)\,,\, J(t^{\prime\prime})\,,\, J(0)\, %
\rangle_0\, \, dt^\prime\, dt^{\prime\prime}\,=\, %
\frac 12 \,K_S(t)\,\,, \, \label{f2_13}
\end{eqnarray}
where $\,K_S(t)\,$ is correlation function of fluctuating power %
$\,S(t)\,$ of equilibrium white noise.

\subsection{Non-Gaussian random walk. %
Characteristic function of displacement}

Let us consider CF (\ref{f2_2}). Its logarithm %
(\ref{f2_5}) for brevity also will be termed CF. %
Expanding CF (\ref{f2_5}) into series over $\,ik\,$ we have %
\begin{eqnarray}
\Delta_t(ik)\,=\, \sum_{m=1}^\infty %
\frac {(ik)^{2m}}{2m\,!}\, D_{2m}(t)\,\, \label{f2_14}
\end{eqnarray}
We took into account that, because of invariance of laws %
of microscopic motion in respect to time reversal, %
equilibrium Brownian motion is spatially symmetric. %
Therefore in (\ref{f2_14}) only even degrees do appear. %
From the probability theory it is known that full number of %
nonzero terms of the series (\ref{f2_14}) is always infinite. %
The only exclusion is ``Gaussian'' case when %
$D_n(t)=0\,$ for all $\,n\geq 3\,$.

According to (\ref{f2_1}), (\ref{f2_5}), (\ref{f2_10}),
\begin{eqnarray}
D_n(t)\,=\, \frac 1t \int_0^\infty %
\langle\, \,v(t_1)\,,\, \dots\,,\, %
v(t_n)\,\rangle_0\, \, dt_1\dots dt_n\, %
\,\, \label{f2_15}
\end{eqnarray}
If observation time $\,t\gg \tau_\mu $\, then %
$D_2(t)=2D\,$, where $\,D\,$ is diffusivity. %
Assume that higher correlators (cumulants) of velocity %
are fast enough decaying at $\,|t_i-t_j|\rightarrow\infty\,$. %
In such case in (\ref{f2_14})-(\ref{f2_15}) limits %
$D_n\equiv \lim_{t\rightarrow\infty} D_n(t)\,$ %
do exist. The quantities $\,D_n\,$ give values of $\,n\,$-order %
poly-spectra at zero frequencies, they can be %
termed also $\,n\,$-order ``non-Gaussian diffusion coefficients''. %
From (\ref{f2_5}), (\ref{f2_14}) we have
\begin{eqnarray}
\Delta_\infty(ik)\,=\, %
\sum_{m=1}^\infty \frac {(ik)^{2m}}{2m\,!}\,D_{2m}\,=\, %
-D\,k^2\,+\,D_4\,\frac {k^4}{24}\, %
-\,\dots\,\, \,\, \label{f2_16}
\end{eqnarray}
If, however, higher-order velocity correlators %
by some reason are slow decaying under separation %
of time arguments, then the diffusion coefficients %
$D_n\,$ ($\,n\geq 4\,$) may become infinite. %
It means that CF (\ref{f2_16}) is non-analytical function %
of $\,ik$\,. In such the case it is convenient to use %
integral representation of CF [12]:
\begin{eqnarray}
\Delta_\infty(ik)\,=\, %
\int_{-\infty}^\infty %
(\cos\, kr\, -\,1)\, \frac {2D}{r^2}\, %
G(r)\, dr \,\,  \label{f2_17}
\end{eqnarray}
The multiplier $\,2D\,$ here is extracted by %
dimensionality reasonings. From (\ref{f2_14})-(\ref{f2_15}) %
it follows that $\,G(-r)=G(r)\,$ and
\begin{eqnarray}
\int_{-\infty}^\infty %
G(r)\, dr\,=\,1 \,\,  \label{f2_18}
\end{eqnarray}
In essence, (\ref{f2_17}) is a kind of Fourier %
transform (taking into account that %
$\Delta\infty(0)=0\,$).

In the probability theory the limit transition %
$\tau_\mu/t \rightarrow 0$\,, analogous to the above %
concerned one, is performed in another, more formal, %
fashion. There $\,t\,$ stays finite and at that %
$\tau_\mu\,$ is turned to zero. The result is %
some ``Brownian'' process with infinitely %
divisible increments, i.e. with independent %
increments [12] \checkmark \cite{c7}.  
The integral representation like (\ref{f2_17})  %
is termed Levy-Khinchin representation. %
A fundamental theorem of the probability theory %
states that the kernel $\,G(r)\,$ always is non-negative, %
$G(r)\geq 0\,$, which just means the infinite divisibility.
In other respects this function is arbitrary (if only %
the integral in (\ref{f2_17}) is converging).

The Gaussian diffusion, when $\,D_n=0\,$ for $\,n\geq 3\,$, %
is single peculiar, degenerated, case corresponding %
to kernel $\,G(r)=\delta(r)\,$. %

Under ``physical'' limit transition, %
when $\,t\,$ grows while $\,\tau_\mu\,$ is fixed, %
the limit of CF (\ref{f2_10}), generally speaking, %
may not correspond to strictly infinitely divisible %
distribution, and $\,G(r)\,$ can be not strictly non-negative. %
However, in various physical problems the increments %
of $\,r(t)$\,, $\,Q(t)\,$, etc., at large $\,t\,$ practically %
possess asymptotic property of infinite divisibility %
(even in presence of slow decaying, non-integrable, %
correlations;\, see example in [15]).

CF (\ref{f2_16}) contains complete information %
about large-scale characteristics of the random walk %
considered in rough macroscopic time scale. %
Corresponding approximate expression for probability %
distribution of $\,r(t)\,$ follows from (\ref{f2_7}) %
after replacement $\,\Theta_t(ik)\rightarrow %
\exp{(t\Delta_\infty(ik))}$\,. %
At $\,t\rightarrow\infty\,$, $\,k\rightarrow 0$\,, %
a dominating role is played by first term of the %
expansion  (\ref{f2_16}), that is diffusion is %
asymptotically Gaussian.

\subsection{Scale invariance of real Brownian %
motion ($\,r^2\propto t\,$)}

The Brownian motion under consideration %
represents a physical process realizing in %
a thermodynamical system and in equilibrium state. %
Like other thermodynamical phenomena, this process %
must not depend (on scales much greater than characteristic %
microscopic scales) on detail structure of %
microscopic interactions. Consequently, it should possess %
some spatial-temporal scale invariance. %
A form of this scale invariance is evidently %
indicated by dimensionality of diffusion coefficient %
$D\,$, the macro-parameter determining large-scale %
properties of Brownian motion (and characteristic %
law of diffusion,\, $\,\langle r^2(t)\rangle_0 =2Dt\,$). %

Another  parameter, what would %
compete with $\,D\,$ in this sense, could appear only %
due to some slow physical processes %
having significant influence on statistics  %
of diffusion.

We assume that there are no such slow processes %
in our system. Then the only additional parameters %
determining (together with $\,D\,$) complete set of %
statistical characteristics of diffusion %
$\{ D_n(t)\}\,\,$ are microscopic quantities describing %
``fast'' interactions of carrier with medium. %
This means that the scale invariance claimed by %
the ``average'' law of diffusion, %
$\,\langle r^2(t)\rangle_0 =2Dt\,$,
governs the whole statistical picture of %
Brownian motion. In other words, it must %
look self-similar when spatial scale %
changes by  $\,\lambda\,$ times while temporal scale %
by $\,\lambda^2\,$ times ($\,r^2\propto t\,$). %

Mathematically this statement reads, %
as one cam see from (\ref{f2_5}), as follows:
\begin{eqnarray}
\lambda^2\, \Delta_{\lambda^2 t} %
\left( \frac {ik}\lambda \right)\,=\, %
\Delta_t(ik)\,\,, \,\, \label{f2_19}
\end{eqnarray}
at sufficiently large $\,t\,$ and small $\,|k|\,$, %
and
\begin{eqnarray}
\lambda^2\, \Delta_\infty %
\left( \frac {ik}\lambda \right)\,=\, %
\Delta_\infty(ik)\,\, \,\, \label{f2_20}
\end{eqnarray}
Below we shall show that such scale invariance %
of Brownian motion (at large scales) realizes %
through flicker fluctuations of diffusivity %
(rate of the motion) with 1/f-type spectrum.

\subsection{1/f-noise as natural attribute of diffusion. %
Spontaneous diffusivity fluctuations. Logarithmically %
decaying correlations}

First, consider relation (\ref{f2_20}). %
Substitution of (\ref{f2_17}) to (\ref{f2_20}), after %
change of variables in the integral we obtain
\begin{eqnarray}
\lambda\,  G(\lambda r)\,=\,G(r)\, \nonumber
\end{eqnarray}
This functional equation has two solutions:
\begin{eqnarray}
G(r)\,=\,\delta(r)\,\,, \,\,\,\,\, %
G(r)\,=\,\frac A{|r|}\,\,, \,\,
\label{f2_21}
\end{eqnarray}
where $\,A=\,$const\,. The first possibility %
leads to CF (\ref{f2_6}), i.e. to ideally Gaussian %
diffusion which has no place in nature. Therefore %
let us consider the second possibility. %

Choice of the second of expressions (\ref{f2_21}) %
result in divergency of integral in (\ref{f2_17}) %
at $\,r\rightarrow 0\,$. This means that the scale %
invariance can not be perfect, that is it must be %
violated at small (microscopic) scales, which is %
obvious from physical viewpoint. Consequently, %
we have to cut-off the integrand, for instance, %
by setting
\begin{eqnarray}
G(r)\,=\,\frac A{|r|+r_0}\,\, \,\, \label{f2_22}
\end{eqnarray}
(below it will be seen that details of cut-off procedure %
are rather  insignificant). Thus we introduce characteristic %
spatial micro-scale $\,r_0\,$. %
It can not be smaller than mean free path (in case of quasi-free %
carriers) 
or mean step under hopping conductivity \checkmark \cite{c8}. %

Inserting (\ref{f2_22}) to (\ref{f2_17}), %
we obtain, at $\,k^2r_0^2\ll 1\,$ (which corresponds to much %
larger scales than $\,r_0\,$), %
\begin{eqnarray}
\Delta_\infty(ik)\, \approx\, -\,Dk^2\, %
A\, \ln\,\frac 1{r_0^2k^2}\,\,\, \label{f2_23}
\end{eqnarray}
\checkmark \cite{c9}\,.\, %
Non-analyticity of this function manifests %
presence of non-integrable long-living higher-order %
correlations of $\,v(t)\,$. %
At that, however, it is unpleasant that function (\ref{f2_22}) %
does not satisfy the necessary condition (\ref{f2_18}), %
and therefore expression (\ref{f2_23}) does not %
identify quadratic term $\,\propto (ik)^2\,$. %
This means that the invariance %
must be destroyed at large scales too, as far as observations of %
the diffusion process take a finite time.

Hence, we have to go back to CF (\ref{f2_5}) %
for finite time and analyse it with the help of %
relation (\ref{f2_19}). With this purpose, let us %
use analogue of the representation (\ref{f2_22}) %
as follows:
\begin{eqnarray}
\Delta_t(ik)\,=\, %
\int_{-\infty}^\infty %
(\cos\, kr\, -\,1)\, \frac {2D}{r^2}\, %
G(r,t)\, dr \,\,, \,  \label{f2_24}\\
\int_{-\infty}^\infty G(r,t)\,dr \,=\,1 \,\,\label{f2_25}
\end{eqnarray}

At $\,t\rightarrow\infty\,$ function $\,G(r,t)\,$ %
should turn into (\ref{f2_22}) (but now with %
coefficient $\,A\,$ depending on $\,t\,$ because of %
condition (\ref{f2_25}). The condition (\ref{f2_25}) %
says that at $\,k\rightarrow 0\,$ %
(and $\,t\gg \tau_\mu\,$) CF (\ref{f2_25}) tends to %
expression $\,-Dk^2\,$, i.e. %
is asymptotically Gaussian. One can verify that %
these reasonings together with (\ref{f2_19}) and (\ref{f2_24}) %
imply the following form of the kernel in representation (\ref{f2_24}):
\begin{eqnarray}
G(r,t)\,=\,\frac {A(t)}{|r|+r_0}\,  %
F\left(\frac {r^2}{4D^\prime t} %
\right)\,\,, \,\,  \label{f2_26}
\end{eqnarray}
where $\,D^\prime\,$ is a constant with same dimensionality %
as $\,D\,$ has,\, $\,A(t)\,$ is determined by the normalization %
condition (\ref{f2_25}),\, function %
$F(z)\,$ satisfies requirements
\begin{eqnarray}
F(0)=1\,\,,\,\,\,\, %
F(z)\rightarrow 0\,\,\, \texttt{at}\,\,\, %
z\rightarrow\infty\,\,, \,\,\,\, %
\int_0^\infty F(z)\,dz\,=\,1\,\, \label{f2_27}
\end{eqnarray}
(they always can be satisfied due to presence of free %
parameters $\,A(t)$\,, $\,D^\prime$\,). %
With use of (\ref{f2_27}) we find from (\ref{f2_25}) that, %
regardless of concrete form of $\,F(z)\,$,
\begin{eqnarray}
A(t)\,=\,\left\{ \int_{-\infty}^\infty %
F\left(\frac {r^2}{4D^\prime t} %
\right)\,\frac {dr}{|r|+r_0}\,\right\}^{-1}\, %
\approx\, \left(\ln\,\frac t{\tau_0} \right)^{-1} %
\,\, \label{f2_28}
\end{eqnarray}
at $\,t\gg \tau_0\,$. Here a microscopic time scale %
has appeared:
\begin{eqnarray}
\tau_0\,=\, \frac {r_0^2}{2D^\prime} %
\,\, \label{f2_29}
\end{eqnarray}

Next, insert (\ref{f2_26}), (\ref{f2_28}) to (\ref{f2_24}) %
and consider first terms of expansion (\ref{f2_24}):
\begin{eqnarray}
\Delta_t(ik)\,=\, -Dk^2 \,+\, \frac 13 \, %
DD^\prime\, t\,A(t)\,-\,\dots\,\,\,, \nonumber\\
D_4(t)\,=\, 8\, DD^\prime\, tA(t)\,\, \, \label{f2_30}
\end{eqnarray}
The second term of series (\ref{f2_30}) contains information %
about diffusivity fluctuations. %
Indeed,  phenomenological relations (\ref{f2_11}), %
(\ref{f2_12}) and definition (\ref{f2_15}) imply  %
\begin{eqnarray}
\frac {d^2}{dt^2}\,\, tD_4(t)\,= \, %
24\,K_D(t)\,\,, \,\,  \, \label{f2_31}
\end{eqnarray}
and then comparison of (\ref{f2_30}) and (\ref{f2_31}) %
yields (for $\,t\gg \tau_0\,$)
\begin{eqnarray}
K_D(t)\,=\, \frac 13\, DD^\prime\, %
\frac {d^2}{dt^2}\,\, t^2\,\left(\ln\, %
\frac t{\tau_0}\right)^{-1}\, \approx\, %
\frac 23\,  DD^\prime\, \left(\ln\, %
\frac t{\tau_0}\right)^{-1}\,\,  \label{f2_32}
\end{eqnarray}
Thus, correlation function of diffusivity %
fluctuations decays by logarithmic law. %
It is obvious, already from dimensionality %
considerations, that corresponding spectrum %
is of 1/f-type.

We see that the flicker fluctuations appear %
as generic property of real Brownian motion 
(which inevitably becomes scale-invariant %
when all microscopic scales %
``are left behind'') \checkmark \cite{c10}. %

Interestingly, these fluctuations %
rather weakly tell on shape of the displacement's probability %
distribution (\ref{f2_7}). As a not complicated analysis %
of (\ref{f2_24}), (\ref{f2_26}) does show, at $\,t\gg \tau_0\,$ %
and $\,k^2r_0^2\ll 1\,$ CF (\ref{f2_24}) has approximately %
universal form
\begin{eqnarray}
\Delta_t(ik)\,=\, \frac {Dk^2} %
{\ln\, \frac t{\tau_0}}\, %
\ln{\left( r_0^2k^2\,+\,c\, %
\frac {\tau_0}{t} \right)}\,\,, \,\label{f2_33}
\end{eqnarray}
where $\,c\,$ is a quantity of order of unit. %
CF (\ref{f2_33}) rather weakly differs from ideally %
Gaussian one, (\ref{f2_6}), although coefficients of %
expansion of (\ref{f2_33}) into series (\ref{f2_14}) %
unboundedly grow with time. %
Correspondingly, difference between distribution  (\ref{f2_7}), %
resulting from (\ref{f2_33}), and Gaussian ``bell'' (\ref{f2_8}) %
is almost unnoticeable. In other words, flicker fluctuations %
$D(t)\,$ do not destroy usual picture of diffusion. %
If we ``let out'' an ensemble of Brownian particles %
to walk from some point, their distribution law %
(evolution of their concentration) is almost %
Gaussian bell (\ref{f2_8}) \checkmark \cite{c11}. 
Moreover, it can be shown %
that fluctuations $\,D(t)\,$ practically %
have no influence on (equilibrium) concentration fluctuations, %
so that the latter can be well described in the %
frameworks of usual ``ideally Gaussian'' model of diffusion.

\subsection{Spectrum of diffusivity fluctuations. %
Quantitative estimates. Origin of the %
``Hooge constant''}

Consider spectral power density of (relative) fluctuations %
of diffusivity,
\begin{eqnarray}
S_{\delta D}(\omega)\,\equiv\, %
\frac {S_D(\omega)}{D^2}\, \equiv\, %
\frac 2{D^2} \int_0^\infty K_D(t)\, %
\cos\,\omega t \,dt\,\, \, \label{f2_34}
\end{eqnarray}
We omit calculation of the Fourier integral. %
The result, in the region of our interest %
($\,\omega \tau_0\ll 1\,$), with good accuracy %
is equal to
\begin{eqnarray}
S_{\delta D}(\omega)\,=\, %
\frac {2D^\prime}{3D}\,  %
\frac {\pi}{|\omega|\, ^
(\,\ln\,|\omega|\tau_0\,)^2} %
\,\, \, \label{f2_35}
\end{eqnarray}

Let us compare this expression with the %
empirical formula (\ref{f1_3}). %
Slow flicker fluctuations $\,D(t)\,$ and fluctuations of %
mobility, $\,\mu(t)\,$, should be connected, as it follows %
from simple physical reasonings, by the %
Einstein relation $\,D(t)=T\mu(t)\,$ (in Sec.3 we shall %
rigorously prove it for steady non-equilibrium state). %
Consequently, $\,S_{\delta D}(\omega)= %
S_{\delta \mu}(\omega)\,$, and we can rewrite %
(\ref{f2_35}) as %
\begin{eqnarray}
S_{\delta \mu}(\omega)\,=\, %
\frac {2\pi\, a(\omega)}{|\omega|}\,\,  %
\,, \,\, \label{f2_36}\\
a(\omega)\,\equiv\, \frac{D^\prime}{3D}\, %
 (\,\ln\,|\omega|\tau_0\,)^{-2}\,=\, %
\frac{r_0^2}{6D\tau_0}\, %
 (\,\ln\,|\omega|\tau_0\,)^{-2}\,\, \nonumber
\end{eqnarray}
Spectrum of relative fluctuations of power, $\,S(t)\,$, %
of summary white noise produced by $\,N\,$ statistically %
independent carriers results from (\ref{f2_35}) after %
division by $\,N$\,:
\begin{eqnarray}
S_{\delta S}(\omega)\,\equiv\, %
S_{\delta D}(\omega)\, \frac 1N\,\, \label{f2_37}
\end{eqnarray}

The parameters $\,r_0\,$ and $\,\tau_0\,$ are micro-scales %
representing lower bounds of the scale invariance of Brownian motion. %
Physically, it seems obvious that in homogeneous medium %
$\,2D\tau_0 \gtrsim r_0^2\,$, i.e. %
$\,D^\prime\lesssim D\,$. Indeed, if already after %
single typical ``free path'' (or ``hop''), %
with length $\,\lambda_0\,$, self-correlation of %
direction of motion vanishes, then $\,r_0\,$ must turn %
to  $\,\lambda_0\,$, and therefore quantity $\,2D\tau_0\,$ %
can not be essentially smaller than %
$\,\lambda_0^2\approx r_0^2\,$\, \checkmark \cite{c12}. 

If temporal invariance takes shape starting just from %
minimal accessible scale %
$\,\sim r_0^2/2D\,$, then  $\,2D\tau_0\approx r_0^2\,\,$ and %
$\,D^\prime\approx D\,$. In this simplest case %
$\,\tau_0\approx\tau_\mu\,$ (with $\,\tau_\mu\,$ %
denoting typical free %
path time, or time between hops from one localized state %
to another). At that, ``width'' %
of function $\,F(r^2/4D^\prime t)\,$ in (\ref{f2_26}) %
coincides, in view of (\ref{f2_27}), with width of %
Gaussian bell (\ref{f2_8}). In this remarkable characteristic case, %
diffusion is invariant to maximal extent, since it is %
described by only two parameters, $\,D\,$ and $\,\tau_0\,$.

Confining ourselves by this situation\,\footnote{\, %
Expectedly, it takes place, in particular, under %
domination of one type of inelastic interaction of carriers %
with thermostat, when all correlations of $\,v(t)\,$ are %
fast decaying during time $\,\sim r_0^2/2D\,$. %
}\,, %
let us set $\,2D\tau_0 =r_0^2\,$ (\,$D^\prime =D$\,) %
and estimate the ``Hooge constant'' $\,a(\omega)\,$ %
(compare (\ref{f2_36}) with (\ref{f1_3})). %

In semiconductors, typical free path time %
$\tau_\mu\sim10^{-12}\,$s\,. Taking %
$\tau_0\sim10^{-12}\,$s\,, at frequency %
$\omega/2\pi =1\,$Hz\, we obtain\, %
$a =(1/3)(\ln\,\omega\tau_0)^{-2} %
\approx 5\cdot 10^{-4}\,$, and %
$\,a\approx 1.2\cdot 10^{-3}\,\,$ at frequency %
$\omega/2\pi =10^4\,$Hz\,, in good agrrement %
with typical experimental values $\,\sim 10^{-3}\,$ %
(see Introduction). Thus our theory gives correct estimate of %
the Hooge constant and, hence, 1/f-noise level in many %
real situations [1].

We see that the ``Hooge constant'', %
$a \approx (1/3)(\ln\,\omega\tau_0)^{-2}\,$, %
is determined by microscopically small time $\,\tau_0\,$ %
and total duration $\,\sim 2\pi/\omega\,$ of 1/f-noise %
observations only. Frequancy dependence of $\,a(\omega)\,$ is %
very weak. But it ensures integrability of spectrum %
(\ref{f2_35})-(\ref{f2_36}) at low frequency %
(as it should be for stationary fluctuations) %
\checkmark \cite{c13}. 

If $\,2D\tau_0\gg r_0^2\,\,$ then, according to (\ref{f2_36}), %
1/f-noise level essentially decreases. This statement agrees with %
observed lowering of noise in strongly doped %
semiconductors [1]. Indeed, there $\,r_0\,$ is determined %
by carriers' scattering  by impurities. %
However, the scale $\,\tau_0\,$ must be determined by %
much more slow interaction with phonons. 
Therefore $\,2D\tau_0\gg r_0^2\,$. %
Let us assume that $\,2D\tau_0\approx \lambda_0^2\,$ with %
$\,\lambda_0\,$ being  mean free path under interaction with phonons %
only (i.e. in pure material). Then in (\ref{f2_36}) %
$\,r_0^2/2D\tau_0\approx r_0^2/\lambda_0^2 \,$, which can be %
written also as $\,r_0^2/2D\tau_0\approx (\mu/\mu_0)^2 \,$, %
in agreement with the empirical Vandamme-Hooge formula %
(see Introduction) \checkmark \cite{c14}. 

\subsection{Microscopic origin of flicker fluctuations}

We demonstrated that if Brownian motion has no inner %
macroscopic scales and therefore possesses natural scale %
invariance, $\,r^2\propto t\,$, then it is accompanied by flicker  %
fluctuations of its intensity (``rate'' of diffusion). %
Already the very premise  implies the statement from %
Introduction that related long-living correlations are manifestation of %
absence of long-living memory in random walks of carriers, %
i.e. these correlations are imaginary (nevertheless, %
one has to describe them statistically like real %
correlations) \checkmark \cite{c15}. 
Both their behavior reflected in  (\ref{f2_32}) and %
logarithmic cut-off of spectrum (\ref{f2_35})-(\ref{f2_37}) %
at very low frequencies are determined by the %
microscale $\,\tau_0\,$ only. In essence, the words about %
``imaginarity'' of flicker fluctuations and absence of %
macroscopic scales are equivalent.

The ``microscopic'' interpretation of obtained results %
enforces to refine some conventional notions. %
The usual view at fluctuations of mobility and conductance %
is that these kinetic characteristics always, at any %
time moment, have exactly definite instant values %
which, however, undergo some slow random perturbations. %
Then one should find mechanisms of the perturbations, %
though such attempts constantly are %
unsuccessful. But there exists quite different point of %
view:\, kinetic characteristics have no certain %
``instant'' values, and just this physical fact %
explains 1/f-noise.

Really, it is not hard to understand that interaction %
of Brownian particle (carrier) with thermostat %
not only impels it to diffusive motion but simultaneously %
causes deviations of its random motion from some %
average regime. The latter can be definitely and quickly %
measured if observing a large ensemble of carriers, %
as is usually done in experiments. %
However, a separate carrier ``knows nothing'' %
about properties of the ensemble and %
is not obligatory at all to display walk with certain %
diffusivity and mobility. The latter concepts %
have a meaning only in respect to a large enough set %
of realizations of random motion, %
while their application to individual random %
(but dynamical) trajectory of carrier is senseless. %
A single separate carrier merely has no %
instant diffusivity and mobility, so that %
one can speak about fluctuations of these kinetic %
quantities in suitable phenomenological language only. %
Both $\,D\,$, $\,\mu\,$ and correlation functions %
$\,K_D(t)\,$, $\,K_\mu(t)$\,, etc., are %
characteristics of ensemble of trajectories. %

Of course, if a thing does not exist %
(as clear dynamical characteristics of motion) %
then it can not fluctuate. %
Therefore we have to complicate the picture. %
The diffusion's property what is empirically %
perceived as flicker fluctuations $\,D(t)\,$ %
(or $\,\mu(t)\,$) is mere consequence of %
temporal uniformity of Brownian motion, %
thermodynamical indifference of the system to %
``rate'' of diffusion. %
Wherever a carrier has occurred at given time moment, %
it each time ``starts from beginning'', and its past %
is of no importance. %
Therefore any deviations from ensemble-average regime %
of motion do not induce a compensation and %
accumulate in time\,\footnote{\, %
Notice that in [8] an attempt was made to introduce to %
a model of diffusion similar non-decaying %
(``residual'') correlations, but they were %
addressed to fluctuation in concentration of %
diffusing quantity. In our model, long correlations %
automatically appear in diffusivity %
fluctuations $\,D(t)\,$.
}\,. %
These deviations, or fluctuations of %
``degree of randomness'' of motion, %
cause neither dynamical nor thermodynamical %
reaction of the system  \checkmark \cite{c16}. 
At the same time, %
they stay in frames of characteristic diffusive %
law $\,r^2\propto t\,$. The result, as we have seen, %
is 1/f-noise.

All that picture, which is slipping away from conventional %
kinetic's ideology, in principle could be described %
in formally rigorous and complete enough way %
in terms of the fourth-order correlators %
(cumulants) (\ref{f1_2}), (\ref{f2_12}), (\ref{f2_13}), %
if we were able to analyze them by methods %
of statistical mechanics. However, such the task %
is extremely complicated, even in case of %
quadratic correlators.

In essence, all the aforesaid concern %
any kinetic quantities, not ``electric'' ones only. %
Thus, likely, we can say that in any system, %
where time-uniform transport of some %
extensive physical quantity takes place, %
there are flicker fluctuations of %
spectral power of random flows (Langevin forces) and, %
in non-equilibrium state, also similar fluctuations of %
irreversible flows. For example, the heat %
transfer must be accompanied by flicker %
fluctuations of thermal conductivity and, %
under temperature gradient, heat flow. %
Another example is given by observed fluctuations %
(with 1/f spectrum) of energy losses in %
quartz resonators (see e.g. [9]) \checkmark \cite{c17}. 

\subsection{General description of %
equilibrium 1/f-noise}

Now, let us take arbitrary two-terminal dissipative electric conductor %
whose leads are shortly connected. %
Consider equilibrium diffusion of charge %
$\,Q(t)\,$ (see (\ref{f2_3})) around this closed circuit. %
Here, also the usual diffusion law is satisfied on average, %
$\,\langle Q^2(t)\rangle_0 =St\,$. %
If transport of charge is determined by microscopically %
scaled processes only, the ``charge Brownian motion''\, %
$\,Q(t)\,$ again must be scale-invariant. %
Obviously, now we can immediately write an expression for %
spectrum, $\,S_{\delta S}(\omega)\,$, of relative fluctuations of %
the white noise power. It is sufficient to make %
in (\ref{f2_35})-(\ref{f2_36}) replacements %
$\,2D\Rightarrow S\,$, $\,r_0 \Rightarrow q_0\,$, where %
$\,q_0\,$ is characteristic microscopic scale of charge transfer, %
while $\,\tau_0\,$ has the same meaning as before. %
Thus, at $\,(\omega\tau_0)^2\ll 1=,$, we find %
\begin{eqnarray}
S_{\delta S}(\omega)\,=\, %
\frac {q_0^2}{3\tau_0 S}\cdot %
\frac {2\pi}{|\omega|}\,  %
(\,\ln\,|\omega|\tau_0\,)^{-2}\,\, \label{f2_38}
\end{eqnarray}

If the conductor represents a homogeneous sample with %
length $\,L\,$, then it is easy to transform (\ref{f2_38}) %
back to (\ref{f2_35})-(\ref{f2_36}) by setting %
$\,q_0 = er_0/L\,$, where $\,e\,$ is electron charge, %
and expressing $\,S\,$ via diffusion coefficient %
and number of (independent) carriers. Here, %
$\,q_0\,$ is charge transported through leads %
in outer circuit (or short-circuit) when an inner %
carrier moves to distance $\,r_0\,$. %

Next, consider, as an example, $\,p-n-$junction %
In this case $\,S=e^2n\,$, where $\,n\,$ is mean number of carriers %
crossing the junction per unit time. If carriers are %
not correlated one with another, then, evidently, %
we can write $\,q_0\approx e\,$. Correspondingly, at %
$\,\tau_0\sim 10^{-7}\,$s\, and $\,\omega/2\pi =1\,$Hz\, %
we obtain from (\ref{f2_38}) estimate
\begin{eqnarray}
S_{\delta S}(\omega)\,\approx\, %
\frac {0.002}{\tau_0 n}\cdot %
\frac {2\pi}{|\omega|}\,\,\, \label{f2_39}
\end{eqnarray}
The scale $\,\tau_0\,$ must be determined by ``life-time'' of %
carrier on the junction, so that %
$\,\tau_0n\gg 1\,$. Relation of $\,\tau_0\,$ to %
physical characteristic times of the system can be %
revealed in a more detail model only.

Of course, in the framework of the presented %
phenomenological statistical theory we can not %
generally expect estimates better than by an order of %
magnitude\,\footnote{\, %
The authors of [1] quite successfully %
describe 1/f-noise in non-uniform semiconductor %
structures on the base of empirical model of mobility %
fluctuations (see Introduction) with $\,a\approx 0.001\,$. %
Interesting exception is presented by diodes with %
the Shottky barrier [1]. %
}\,. %

\section{Current 1/f-noise in steady %
non-equilibrium state and nonlinear %
fluctuation-dissipation relations (FDR)}

\subsection{Characteristic function of transported %
charge. Cubic FDR}

 The use of mathematical formalism of non-Gaussian %
random processes gave us possibility to consider 1/f-noise %
in spite of leaving it in equilibrium state. %
Under switching on an electric field, fluctuations  of %
diffusivities of carriers, $\,D(t)$\,, %
transform into fluctuations of their mobilities, %
$\,\mu(t)$\,, and thus of conductance, $\,g(t)\,$, %
and become source of current 1/f-noise. %
For correct analysis of this non-equilibrium state %
we shall apply the nonlinear %
fluctuation-dissipation relations (FDR) %
described in [16]-[21] \checkmark \cite{c18}. 

Let us return to above mentioned tw-terminal conductor. %
Let before time moment $\,t=0\,$ it is in equilibrium with %
its surroundings (and short-circuited) but after %
$\,t=0\,$ is subject to a given constant voltage %
$\,x(t)=\,$const\,. In such the case, as shown in [16,17] %
(see also [18,21]), due to the time reversibility of %
microscopic dynamics, the following exact generating FDR %
takes place\,\footnote{\, %
This formula concerns the conductor itself and all the %
surroundings in contact with it, assuming their mutual %
thermodynamical equilibrium at common temperature %
$\,T\,$ before an external perturbation (electric voltage). %
It is asumed also that there is no external magnetic field. %
}\,: %
\begin{eqnarray}
\left\langle\,\exp{\left\{ \int_0^t \left[\, %
iu(t^\prime)\,-\,\frac xT\,\right]\, %
J(t^\prime)\, dt^\prime \right\}}\, %
\right\rangle_x\,=\, \nonumber\\ \,=\, %
\left\langle\,\exp{\left\{ - \int_0^t %
iu(t-t^\prime)\, J(t^\prime)\, %
dt^\prime \right\}}\, \right\rangle_x\, %
\,, \,\, \label{f3_1}
\end{eqnarray}
where $\,u(t)\,$ is arbitrary probe function. The %
subscript ``x'' reminds of non-equilibrium character of %
fluctuations. Taking here %
$\,u(t)=u=\,$const\,, introduce CF
\begin{eqnarray}
\Delta_t(iu\,|\,x)\,\equiv\, %
\frac 1t\,\ln\, \langle\,\exp{\{\, %
iu \int_0^t J(t^\prime)\, dt^\prime\,\}}\, %
\rangle_x\,\,\, \label{f3_2}
\end{eqnarray}
In equilibrium case it coincides with (\ref{f2_5}). %
From (\ref{f3_1}) it follows that
\begin{eqnarray}
\Delta_t(iu\,-\,x/T\,|\,x)\,=\, %
\Delta_t(-iu\,|\,x)\, \,\,\, \label{f3_3}
\end{eqnarray}
Expansion of (\ref{f3_3}) into series over $\,iu\,$ implies %
an infinite chain of FDR connecting average value, %
$\,\langle Q(t)\rangle_x$\,,  of charge transported during %
observation time $\,t\,$, and various cumulants of %
fluctuations $\,J(t)\,$ and $\,Q(t)$\,. %
With the  help of these FDR, it can be shown, in particular, that
\begin{eqnarray}
\langle \,Q(t)\,\rangle_x\,=\, %
\sum_{m=1}^\infty c_m\, \left( %
\frac xT\right)^{2m-1}\, %
\langle \,Q^{(2m)}(t)\,\rangle_x\,\,, %
\,\, \label{f3_4}
\end{eqnarray}
where
\begin{eqnarray}
\langle \,Q^{(n)}(t)\,\rangle_x\,\equiv\, %
\int_0^t \langle \,J(t_1)\,,\,\dots\,,\, %
J(t_n)\,\rangle_x\, \,dt_1 \dots dt_n\,\, %
\, \label{f3_5}
\end{eqnarray}
are $\,n\,$-order cumulants of transported charge, %
and the numbers $\,c_n\,$ are defined by formulae
\[
\sum_{m=1}^\infty c_m\, z^{2m-1}\,=\, %
\tanh\, \frac z2\,\,, \,\,\,\,\, %
c_1=\frac 12\,,\,\, %
c_2=-\frac 1{24}\,,\,\, %
c_3=\frac 1{240}\,,\,\dots \,
\]
Differentiation of (\ref{f3_4}) in %
respect to $\,t\,$ leads to fluctuational representation of %
average current:
\begin{eqnarray}
\langle \,J(t)\,\rangle_x\,=\, %
\frac xT \int_0^t \langle \,J(t)\,,\, %
J(t^\prime)\,\rangle_x\, \,dt^\prime\, -\,  %
\label{f3_6} \\ \,-\, %
\frac 16 \left(\frac xT\right)^3 %
\int_0^t \langle \,J(t),J(t_1),J(t_2), %
J(t_3)\,\rangle_x\, \,dt_1 dt_2 dt_3\, %
+\,\dots\, \nonumber
\end{eqnarray}

Consider weakly non-equilibrium state and expand %
quantities in first row of (\ref{f3_6}) into %
series over $\,x\,$:
\begin{eqnarray}
\langle \,J(t)\,\rangle_x\,=\, %
g_1(t)\,x \,+\, g_3(t)\, x^3\,+\, %
\dots\,\,, \,\label{f3_7}\\
\langle \,J(t),J(t^\prime)\,\rangle_x\,=\,  %
\langle \,J(t),J(t^\prime)\,\rangle_0\, +\, %
x^2\, R(t,t^\prime)\,+\,\dots\,
\label{f3_8}
\end{eqnarray}
(thus, for simplicity, we assumed the conductor %
electrically symmetric). %
Extracting in (\ref{f3_6}) linear terms, in respect to %
$\,x\,$, we find
\begin{eqnarray}
g_1(t)\,= \,\frac 1T \int_0^t %
\langle \,J(t),J(t^\prime)\,\rangle_0\,  %
\,dt^\prime\,\,, \, \label{f3_9}
\end{eqnarray}
that is usual fluctuation-dissipation theorem (FDT). %
Extraction of cubic terms yields ``cubic FDT'':
\begin{eqnarray}
g_3(t)\,= \, \frac 1T \int_0^t %
R(t,t^\prime)\,dt^\prime\,-\, %
\label{f3_10} \\ \,-\, %
\frac 1{6T^3} %
\int_0^t \langle \,J(t),J(t_1),J(t_2), %
J(t_3)\,\rangle_0\, \,dt_1 dt_2 dt_3\, \nonumber
\end{eqnarray}
Notice that similar quadratic and cubic FDR %
were in part investigated by Efremov [9] and %
Stratonovich [20].

\subsection{Weakly non-equilibrium state. %
Statistical expression for correlation function %
of conductance fluctuations}

Let us consider relation (\ref{f3_10}) at %
$\,t\gg \tau_\mu\,$, where $\,\tau_\mu\,$ %
is correlation time of equilibrium current noise %
in short-circuited conductor. %
The function $\,g_3(t)\,$, representing cubic %
part of average current response, covers %
contributions from both ``fast'' electric %
processes (responsible for conductivity in itself) %
and comparatively slow thermal processes, %
including changeû of conductance because of %
Joule heat. Hence, generally %
$\,g_3(t)\,$, in contrast to $\,g_1(t)\,$, is not constant %
at $\,t\gg \tau_\mu\,$ but can contain %
slow long ``tail''. Moreover, this tail may be %
non-stationary.

The function $\,R(t,t^\prime)\,$, as one can see from %
(\ref{f3_8}), describes ``excess'' fluctuations of %
current in weakly non-equilibrium state, i.e. in %
quadratic regime in respect to voltage (and mean current). %
Hence, in particular, $\,R(t,t^\prime)\,$ contains %
information about flicker fluctuations of current. %
From (\ref{f3_10}) and (\ref{f2_18}) it follows that %
low-frequency (flicker) current fluctuations in the %
quadratic regime are determined, first, by  equilibrium %
fluctuations of white noise power and, second, by slow %
or non-stationary thermal processes (caused by %
Joule heating). Notice that the second source, %
possibly, is related to the so-called %
``temperature'' 1/f-noise what is sometimes %
observed in metals (see Introduction) \checkmark \cite{c19}.

Here, we are interested in the first source %
of current 1/f-noise having thermodynamically %
equilibrium nature. %
Therefore we shall neglect non-equilibrium %
thermal processes, i.e. suppose that at %
$\,t> \tau_\mu\,\,$ the system tends to strictly %
stationary state. In such the case $\,g_3(t)\,$ %
turns to constant at $\,t> \tau_\mu\,$. %
Besides, at $\,t\,,t^\prime \,\gg \tau_\mu\,\,$ %
the function $\,R(t,t^\prime)\,\,$ becomes a function %
of the time difference only:\, %
$\,R(t,t^\prime) \equiv  R(t-t^\prime)\,$. %
Then we can exploit he phenomenological concept %
of conductance fluctuations and write %
(at $\,t\,,t^\prime \,\gg \tau_\mu\,$ and, clearly,\, %
$\,|t-t^\prime| \,\gg \tau_\mu\,$) %
\begin{eqnarray}
R(t,t^\prime) \,=\, R(t-t^\prime)\, %
=\, K_g(t-t^\prime)\,\,, \, \label{f3_11}
\end{eqnarray}
where $\,\,K_g(t-t^\prime)\,\,$ is correlation function %
of these fluctuations\,\footnote{\, %
Notice that use of model of conductance fluctuations %
automatically presumes stationarity. In other words, %
contribution of non-stationary processes (if any) %
to current 1/f-noise can not be described in this %
model surely exploited in the literature. %
\checkmark \cite{c20} %
}.  
Next, differentiate (\ref{f3_10}) in respect to $\,t\,$. %
Since at that the left side of (\ref{f3_10}) %
vanishes \checkmark \cite{c21}, from %
(\ref{f3_10})-(\ref{f3_11}) and (\ref{f2_13}) %
we obtain
\begin{eqnarray}
K_g(t)\,= \, \frac 1{2T^2} \int_0^t %
\! \int_0^t    %
\langle \,J(t),J(t^\prime),J(t^{\prime\prime}), %
J(0)\,\rangle_0\, \,dt^\prime %
dt^{\prime\prime}\,\,, \label{f3_12}\\ %
K_g(t)\,=\,K_S(t)/4T^2\,\, \,\label{f3_13}
\end{eqnarray}

Thus, correlation function of conductance %
fluctuations (caused by equilibrium processes) %
is expressed through fourth-order cumulant of %
equilibrium current fluctuations. %
In essence, formula (\ref{f3_12}) must be %
treated as statistical definition of %
$K_g(t)\,$, since the fourth equilibrium %
cumulant of current always can be represented %
in rigorous terms of statistical mechanics. %

Because of (\ref{f3_12}) and the Nyquist formula, %
$S=2Tg\,$, spectra of relative flicker fluctuations %
$g(t)\,$ and $\,S(t)\,$ are identical and (in weakly %
non-equilibrium steady state) both coincide %
with spectrum of relative fluctuations of current. %
From (\ref{f2_38}) and (\ref{f3_12}), in the %
quadratic regime, we obtain
\begin{eqnarray}
S_{J}(\omega)\,=\, %
\overline{J}^2\cdot \frac %
{q_0^2}{3\tau_0 S}\cdot %
\frac {2\pi}{|\omega|}\, %
 (\,\ln\,|\omega|\tau_0\,)^{-2}\,\,, %
\, \label{f3_14}
\end{eqnarray}
where $\,\overline{J}=gx\,$. %
In case of homogeneous conductor in the %
approximation of independent carriers %
formulae (\ref{f2_35}), (\ref{f2_37}) and %
(\ref{f3_12}) yield
\begin{eqnarray}
S_{J}(\omega)\,=\, %
\overline{J}^2\cdot  %
\frac {2\pi\,a(\omega)}{N\,|\omega|}\,\,  %
\,, \,\, \label{f3_15} \,\,\,\,
a(\omega)\,=\, \frac{r_0^2}{6D\tau_0}\, %
 (\,\ln\,|\omega|\tau_0\,)^{-2}\,\, %
\end{eqnarray}

As was noticed above, this result is in %
rather satisfactory quantitative agreement with %
general empirical formula (\ref{fh}). %
Since we based om ``first principles'' only, %
such result can be qualified as important evidence %
in favor of our approach.

\subsection{Nonlinear conductor. Current %
1/f-noise in non-Ohmic regime}

Our analysis is valid for any (nonlinear) conductor, %
but in the Ohmic (quadratic in respect to current) %
regime. For description of non-Ohmic regime it is %
necessary to take into account non-Gaussianity of white %
noise in itself which, according to the FDR [17,21], %
is in close relationships with dissipative non-linearity. %
It is natural to suppose that character of these %
relationships is not affected by slow flicker fluctuations. %
In the corresponding statistical model all kinetic %
parameters of white noise fluctuate %
in coordination with each other. This can be %
mathematically formulated (for steady state at %
$t\gg \tau_\mu\,$) as follows %
(compare with (\ref{f2_9})): %
\begin{eqnarray}
\Delta_t(iu|x)\,=\, \frac 1t\, %
\ln\, \langle \exp{\{ \int_0^t %
\xi(t^\prime)\, S(iu|x)\,dt^\prime\, %
\}} \rangle_x^\prime\,\,,
\,\, \label{f3_16}
\end{eqnarray}
where\, $\,\int_0^t \xi(t^\prime)\, dt^\prime\,\,$ %
is already considered process of scale-invariant %
diffusivity fluctuations, with %
$\,\xi(t) =S(t)/S\,$, and %
$\,S(iu|x)\,$ is CF of white noise \checkmark \cite{c22}. 
The latter, according to (\ref{f3_3}),  should %
satisfy similar relation
\begin{eqnarray}
S(iu\,-\,x/T\,|\,x)\,=\, %
S(-iu\,|\,x)\,\,  \, \label{f3_17}
\end{eqnarray}
Of course, in general $\,S$\,, as well as %
$\,r_0\,$ and $\,\tau_0\,$, is dependent on $\,x\,$. %

Let us write the white noise CF as series
\begin{eqnarray}
S(iu|x)\,=\,iu\,\overline{J}(x) %
\,+\,\frac {(iu)^2}2 \, S(x)\, %
+\,\dots \,\,, \, \label{f3_18}
\end{eqnarray}
where $\,\overline{J}(x)\,$ is average current, %
and $\,S=S(x)\,$ spectral power of noise. %
Combining (\ref{f3_16}) with (\ref{f3_18}), %
it is easy to obtain, for current flicker %
fluctuations,
\begin{eqnarray}
K_J(t)\,=\,\,\overline{J}^2(x)\, %
\frac {2q_0^2(x)}{3\tau_0 S(x)}\, %
\left(\ln\,\frac t{\tau_0}\right)^{-1}\,
\,\, \label{f3_19}
\end{eqnarray}
and, correspondingly,
\begin{eqnarray}
S_{J}(\omega)\,=\, %
\frac {\overline{J}^2(x)\,q_0^2(x)} %
{3\tau_0 S(x)}\cdot %
\frac {2\pi}{|\omega|}\, %
\label{f3_20} %
(\,\ln\,|\omega|\tau_0\,)^{-2}\,\, %
\end{eqnarray}

Thus, spectrum of current fluctuations %
is presented, as before, by expression like %
(\ref{f2_38}), but now with taking into account %
possible dependences on $\,x\,$. %
At that, FDR (\ref{f3_17}) dictates definite, %
may be rigid, relations between %
$\,\overline{J}(x)\,$ and $\,S(x)\,$.

For example, consider again an ideal %
semiconductor diode ($\,p-n-$junction) in %
non-Ohmic regime, or shot noise regime. %
In this case, as is known, %
\begin{eqnarray}
S(x)\,=\,e\, |\overline{J}(x)|\, %
\,\, \label{f3_21}
\end{eqnarray}
at $\,\,e|x|\gg T\,$. %
Assuming that $\,q_0=e\,$ and $\,\tau_0\,$ do not %
change, we find from (\ref{f3_20})-(\ref{f3_21}) %
that
\begin{eqnarray}
S_J(\omega)\,\approx\, \frac %
{e\,|\overline{J}(x)|}{\tau_0}\cdot %
\frac {2\pi\,a(\omega)}{|\omega|}\, %
\,\, \label{f3_22}
\end{eqnarray}
Hence, in nonlinear regime intensity %
of current flicker noise grows %
proportionally to mean current, %
that is merely to number of carriers %
crossing the junction per unit time (see [1]) %
\checkmark \cite{c23}.

\section{Conclusion}

Let us resume main aspects and results of the present work.

{\bf 1}\,. The experimental situation around 1/f-noise %
does not fit in frames of traditional notions. %
The matter is in confusion between concepts %
concerning statistical properties of ensembles and %
concepts concerning individual charge carrier random walks. %
The confusion is unconsciously provoked by use of traditional %
Gaussian model of diffusion.

{\bf 2}\,. Our principal result is that 1/f-noise %
can be explained without attraction of special %
physical mechanisms, merely as generic property of %
random Brownian motion of charge carriers, i.e. %
the well known process involved in most of %
electric phenomena.

{\bf 3}\,. We constructed a simple novel model of %
Brownian motion, exploiting only the same physical %
premises as the standard model do, but refusing %
assumption about Gaussian character of statistics %
of real Brownian motion, since this is physically %
senseless idealization \checkmark \cite{c24}. 
Thus our theory is more adequate (though, of course, %
may require further improvements).

{\bf 4}\,. The constructed theory inevitably leads to %
conclusion that real Brownian motion always %
possesses fluctuations of ``rate of diffusion'' %
with 1/f-type spectrum.

In the ensemble language one can say that diffusion %
coefficient and spectral power of ``white'' %
electric noise undergo fluctuations with 1/f-type spectrum. %
In non-equilibrium states these fluctuations manifest %
themselves in mobilities of carers, conductance and current.

{\bf 5}\,. Random motion of separate carrier (and, generally, %
any concrete Brownian trajectory) has no certain diffusivity %
(diffusion coefficient). What is for the ``rate of diffusion'', %
its spontaneous (thermodynamic) fluctuations have no %
characteristic time scale, since make not back impact upon %
dynamical picture of motion or thermodynamical state %
of the system.

1/f-noise is just result of absence of long-living %
correlations in mechanisms of random motion of 
carriers (and has no relation to some macroscopically %
large relaxation times). Therefore, spectral contents %
of 1/f-noise is indifferent to system's sizes.

{\bf 6}\,. Intensity of 1/f-noise is determined %
by microscopic scales only. In the constructed model, %
there are only two such parameters %
which indicate spatial and temporal lower %
bounds of scale invariance of Brownian motion.

{\bf 7}\,. The resulting estimate of 1/f-noise %
intensity is in satisfactory agreement with experiments. %
At that, origin of the small empirical ``Hooge constant'' %
is revealed.

{\bf 8}\,. A relevant microscopic information  %
about 1/f-noise in stationary states can be %
obtained, in principle, from fourth-order %
equilibrium correlators (cumulants), with use of %
rigorous methods of statistical mechanics. %
This is conceptually quite new problem of %
theoretical physics.

\appendix

\section*{Appendix}

Let us show now in the spectral representation, %
i.e. in the frequency domain, that fluctuations of %
power of non-Gaussian equilibrium white noise %
and corresponding conductance fluctuations %
possess 1/f-type spectrum. %

Let $\,J(\omega)\,$ be Fourier transform of current %
fluctuations. According to the Wiener-Khinchin theorem, %
\begin{eqnarray}
\langle\, J(\omega),J(\omega^\prime)\, %
\rangle_0\,=\, 2\pi\,S\, %
\delta(\omega +\omega^\prime)\,\,, \, \nonumber
\end{eqnarray}
where $\,S=\,$const\, for white noise. %
From here it follows that
\begin{eqnarray}
\langle\, J(\lambda^2\omega), %
J(\lambda^2\omega^\prime)\, %
\rangle_0\,=\, 2\pi\,S\, %
\delta(\lambda^2(\omega +\omega^\prime))\, %
=\,\, \nonumber\\ \,=\, %
2\pi\, \frac S{\lambda^2}\,\, %
\delta(\omega +\omega^\prime)\,=\, %
\left\langle\, \frac {J(\omega)}\lambda\,, %
\frac {J(\omega^\prime)}\lambda\, %
\right\rangle_0\, \, \nonumber
\end{eqnarray}
Hence,
\begin{eqnarray}
\lambda\, J(\lambda^2\omega)\,\propto\, %
J(\omega)\,\,, \,\,  \label{a1}
\end{eqnarray}
where symbol $\,\propto\,$ means identity in %
statistical sense. If current fluctuations %
have no coupling with some slow processes, %
then the scale invariance expressed by formula %
(\ref{a1}) must extend to the whole statistics %
of the fluctuations. Taking this in mind, %
consider fourth-order spectral cumulant:
\begin{eqnarray}
\langle\, J(\omega_1),J(\omega_2), %
J(\omega_3),J(\omega_4) \,\rangle_0\,=\, %
2\pi\,S_4(\omega_1,\omega_2,\omega_3)\, %
\delta(\omega_1 +\omega_2+ %
\omega_3 +\omega_4)\,\,, \,  \label{a2}
\end{eqnarray}
where $\,S_4\,$ is tri-spectrum. %
According to the analogue of (\ref{f2_11}) %
for current,
\begin{eqnarray}
S_4(\omega_1,\omega_2,\omega_3)\,=\, %
S_S(\omega_1 +\omega_2)\,+\, %
S_S(\omega_1 +\omega_3)\,+\, %
S_S(\omega_2 +\omega_3)\,\,  \label{a3}
\end{eqnarray}
Replacing in (\ref{a2}) $\,J(\omega)\,$ by %
$\,\lambda\, J(\lambda^2\omega)\,\,$ %
and combining the result with identity (\ref{a1}) and %
expression (\ref{a2}) itself, it is easy to deduce that %
\begin{eqnarray}
\lambda^2\, S_4(\lambda^2\omega_1, %
\lambda^2\omega_2, \lambda^2\omega_3)\,=\, %
S_4(\omega_1,\omega_2,\omega_3)\,\,, \,\,\nonumber\\
\lambda^2\, S_S(\lambda^2\omega)\,=\, %
S_S(\omega)\,\, \,\,\nonumber
\end{eqnarray}
From here we find
\begin{eqnarray}
S_S(\omega)\,=\, \frac %
{\texttt{const}}{|\omega|}\,\,, \,\label{a4}
\end{eqnarray}
that is the power and conductance fluctuations %
have 1/f-type low-frequency spectrum %
(solution $\,S_S(\omega)\propto \delta(\omega)\,$ %
is not appropriate since $\,S_4\,$ must be a smooth function). %
Our above results differ from this more formal one %
only by slight distortion of ideal scale invariance.




\begin{thebibliography}{58}
\bibitem{c1}
To be more precise, in electrolytes $\,a\,$ may achieve as large %
values as $\,a\sim 10\,$ \cite{bk3}. %

\,\,\,\, Useful information about temperature dependence of $a$ %
in semiconductors can be found in \cite{fnh}.

\,\,\,
\bibitem{c2}
Unceasing search of ``fluctuation modes'', %
or ``fluctuators'', to form 1/-type spectru, %
continues up to now. But, strangely, its %
otioseness does not push to simple idea that %
``fluctuators'' are merely unique trajectories %
(``fates'') of charge carriers.

\,\,\,
\bibitem{c3}
Highly dubious hypothesis ignoring the fact that %
energy (number of phonons) in any partiular vibration (phonon) %
mode possesses rather short (at least, not giant) %
relaxation, or thermalization, time (while fluctuations in %
summary energy density of different phonon modes are  %
the same as the temperature fluctuations already shown to fail %
as a candidate for source of 1/f-noise).

\,\,\,\, In reality, as shown in \cite{i3}, observed %
1/f-noise in (equilibrium) phonon system %
represents fluctuations of rates of relaxation %
(relaxation times) of phonon modes %
(see also discussion in \cite{eiphg}).

\,\,\,
\bibitem{c4}
From viewpoint of the standard naive logics (see e.g. %
\cite{w}), this paradox enforces to search  %
slow ``fluctuators'' what randomly modulate conductance %
with magnitude inversely proportional to $\,\sqrt{N}\,$, %
as if number of fluctuators was proportional to $\,N\,$. %
Rather difficult task (ensuring employment of the logicians %
for many years in future)!

\,\,\,\, More adequate logics should take into account that %
slowly decaying statistical correlations in general do not %
presume slowly decaying cause-and-consequence correlations %
(long-living memory). In opposite, long statistical correlations %
(or even not decaying ones) can arise from indifference of %
system to pre-history of transport processes in it (i.e. just from %
absence of long memory)! For explanations see %
e.g. \cite{i2,yuk} and introductory and/or discussion-conclusion %
sections in other our works, \cite{i1} - \cite{p1208}. %

\,\,\,\, In brief, if charge conduction is indifferent to amount of already %
transortted charge $\,Q(t)\,$, then it is unable to ensure %
well certain value of time-averaged current $\,Q(t)/t\,$, which just %
means presence of arbitrary long statistical correlations between %
time-local current values.

\,\,\,\, That is what must be expected from rigorous statistical %
mechanics \cite{kr}, in place of its crucial roughenings %
in conventional kinetics (pathological simplifications %
``mutilating'' the very way of thinking of theoreticians). %

\,\,\,
\bibitem{c8}
To be more precise, in general there are more %
than one characteristic ``free path'' %
lengths, and apparently $\,r_0\,$ can not %
be smaller than minimum of them. %
For example (touched in Introduction and below), %
if energy relaxation of carriers is more slow than %
velocity relaxation, then we have at least two %
different free paths, ``elastic'' and ``inelastic''. %
Then it seems natural to equate  $\,r_0\,$ to %
shortest of them (``elastic''), while associating\, %
$\,\sqrt{2D\tau_0}\,$ with the longest (``inelastic''), %
so that $\,r_0^2/2D\tau_0 =D^\prime/D \ll 1\,$, %
i.e. diffusivity 1/f-noise is much weaker than its %
``seed'' refernce level corresponding to $\,D^\prime/D = 1\,$.

\,\,\,\, The opposite situations, when %
$\,r_0^2/2D\tau_0 =D^\prime/D \gg 1\,$, are possible too. %
Some  examples can be found in \cite{bk3,pr2,p1008}. %
Besides, recollect that in metals notably below Debye %
temperature, vice versa, velocity (direction) relaxation is %
slower than energy relaxation.

\,\,\,
\bibitem{c11}
As was shown in \cite{pr2}, probability weight of %
non-Gaussian (more or less long) tails of the $\,r(t)\,$'s %
distribution typically is around 0.03\,. %
Nevertheless, this means that probabilities of large deviations %
of $\,r(t)\,$ from typical diffusion rate %
(i.e. probabilities of $\,|r(t)|\gg \sqrt{2Dt}\,$) %
always are very mich greater than in Gaussian model %
(see \cite{p0802,yuk,p1209,bkn}).

\,\,\,\, Of course, statistics of large deviations %
is sensitive to details of the function $\,F(z)\,$, %
especially if it decays by a power law, e.g. %
$\, F(z) =\,1/(1+z/\beta)^{1+\beta}\,$ (we took into account %
conditions (\ref{f2_27})). At that, however, to satisfy actual %
finiteness of all $\,r(t)\,$'s cumulants, we should  %
multiply such function of $\,z=r^2/4D^\prime t\,$ %
b a cut-off factor $\,F^\prime(r^2/v_0^2t^2)\,$, where %
$\,F^\prime(z^\prime )\,$ is fast decreasing function of %
$\,z^\prime r^2/v_0^2t^2\,$ and $\,v_0\,$ is characteristic %
velocity of carriers (e.g. thermal or Fermi velocity). %
Then probability distribution of %
$\,r(t)\,$ asymptotically (at $\,v_0^2t^2/D^\prime t\, %
\rightarrow\infty\,$) obeys scale invariance $\,r^2\propto t\,$ %
but far tails of this distribution and its higher-order %
cimilant significantly deviate from $\,r^2\propto t\,$ %
(therefore, in particular, the ``noise of 1/f-noise'' %
\cite{pr2} can be of 1/f$^\gamma\,$-type with $\,\gamma >1\,$). %
A similar picture arises in statistical-mechanical %
theory of ``molecular random walks'' %
\cite{p12,p0802,p0803,tmf,p1209,bkn}.

\,\,\,
\bibitem{c12}
Of course, that are not absolutely convincing reasonings, %
and one can suggest contrary instances to them. %
For example, in metals, due to the Fermi statistics of %
carriers, even under simple inelastic scattering %
(above the Debye temperature) %
$\,r_0^2\approx \lambda_0^2\gg 2D\tau_0\approx 2D\tau_\mu\,$, %
so that the Hooge constant rises nearly $\,T_F/T\,$ %
times as compared to the ``seed'' level \cite{bk3} %
(as if only $\,\sim (T/T_F)\,N\,$ of %
total number $\,N\,$ of conduction electrons, - from vicinity %
of Fermi surface, - effectively contribute to 1/f-noise).

\,\,\,\, On some of other possible variants of %
relationships between characteristics %
micro-scales see \cite{bk3,pr2,i2,p12,p0802,kmg,p1008}. %
Unfortunately, exact equations of microscopic %
Hamiltoniam model of electron's random walk in %
phonon field \cite{eiphg} still %
are not analysed up to quantitative estimates %
to compare with the commented phenomenology.

\,\,\,
\bibitem{c13}
Though, strictly speaking, this integrability %
is not consequence of stationarity %
in its physical sense. %
In other words, 1/f-noise can look formally %
non-stationary even in thermodynamically equilibrium %
(all the more, in steady non-equilibrium) systems!
This important theoretical fact was demonstrated and %
explained already in \cite{i1} (se also \cite{i2,p12}).

\,\,\,
\bibitem{c15}
This seeming paradox is main (practically fatal!) %
obstacle in ways to adequate theoretical perception %
of 1/f-noise.

\,\,\,\, Though, it is trivial logical truth that %
(long) statistical correlations %
can take place without (long) physical (cause-and-consequence) %
correlations beyond them. Moreover, forgetting %
of the past (history of a transport process) means impossibility %
to keep the future (fluctuations in transport rate) %
predictable and controllable.  %
Therefore short-memory transport process inevitably acquires %
scaleless long-term fluctuations (and thus statistical correlations) %
of its instant rate.

\,\,\,\, Nevertheless, physicists (even theoreticians) %
accept for a fact the prejudice %
that physical independence is %
the same as statistical independence. %
%
Consequently, they continue, like thirty years ago, %
to search sources of 1/f-noise observed %
in transport processes (and other irreversible and %
transition processes) everywhere except randomness of %
these processes in itself\,! %

\,\,\,\, Notice that measurements of effective ``Hooge %
constant'' (``Hooge parameter'') in very different systems, - %
including modern small-size diamond, grathene and %
carbon nanotube based devices (see, for example, \cite{dsg}, %
\cite{gsg} and \cite{cnt,cnt1}, respectively, and %
references in \cite{p1008}), - %
usually show values from rather narrow interval %
$\,10^{-4}\lesssim a\lesssim 10^{-1}\,$, in spite of %
giant variety of conditions and regimes of charge transport %
under investigations. %
But, strikingly, this fact does not enforces investigators %
to assume that mysterious source of 1/f-noise %
is neither charge traps nor structural defects or impurities %
(or even phonon energy fluctuations, - see discussions %
in \cite{i3,eiphg}). %
That is what the Koz'ma Pritkov's joke is about %
(see epigraph above).

\,\,\,\, In reality, of course,  the enumerated factors, along %
with many others, - e.g. Coulomb interaction between  carriers %
and related screening and space charge effects, - all can essentially %
influence 1/f-noise, but this does not %
mean that they are its sources (similarly, they %
can influence white noise although not being its sources). %

\,\,\,\, Hence, the actual problem waiting for theoreticians %
is development of dynamical (i.e. statistical-mechanical) %
models of systems under practical interest (thus continuing our %
first steps made in \cite{bk3}, \cite{i1} - \cite{eiphg}, \cite{bkn}). %

\,\,\,\, The phenomenology we are commenting now %
is too primitive from viewpoint of such  %
complicated inhomogeneous structures as  %
mentioned various field-effect transistors %
\cite{dsg,gsg,cnt,cnt1}. %
Nevertheless, it can be notably helpful %
in interpretation of experimental data %
(see comment \cite{c23} below).

\,\,\,\,
\bibitem{c16}
``Rate of diffusion'' by its sense is a time-nonlocal quantity %
smeared over time intervals much longer than system's %
memory for diffusion. That is why this quantity %
has no definite instant values, is free of system's back reaction, %
and fluctuates with no certain time scale.

\,\,\,\,
\bibitem{c18}
In English, see also papers \cite{fds} plus \cite{p0501} - \cite{bkn} %
and other our works on the nonlinear FDR, %
or ``generalized FDR'', referenced in \cite{p0501} - \cite{bkn}.

\,\,\,
\bibitem{c19}
See additionally \cite{bk3} and references therein. %
Some usefull information can be found also in \cite{w}.

\,\,\,
\bibitem{c20}
Though, see above comment \cite{c13}. %
Here, we again have to distinguish between %
formal and physical stationarities, %
since they are not equivalent, %
and integral in (\ref{f3_12}) can %
be unboundedly growing function of $\,t\,$ %
although being characteristics of equilibrium state. %
This means \cite{i1} that result of conductance measurement %
depends on its duration  $\,t\,$ but in no way on time when it %
begins, and at that does not improve when $\,t\,$ increases %
(in opposite, longer time averaging makes it worse, %
since increases its mean-square relative deviation %
from ensemble average\,!). %
In such case (\ref{f3_12}) must be treated like %
the structure function of non-stationary random process %
(for examples see  \cite{i1,p12}), and relations between %
theoretical expressions and experimental procedures need %
in more careful analysis. %

\,\,\,
\bibitem{c21}
To be precise, it anyway decreases with time faster %
than $\,\propto 1/t\,$, while $\,K_g(t)\,$ by its sense %
decreases slower than $\,\propto 1/t\,$ %
(if not increases, - see above).

\,\,\,
\bibitem{c22}
Clearly, the CF $\,S(iu|x)\,$ here is the same %
as $\,\Delta_\infty(iu|x)\,$ in absence of low-frequency %
flicker fluctuations which are separated %
into factor $\,\xi(t)\,$. At that, the latter can be treated as %
derivative of randomly non-uniform ``inner time'' of
the transport process, %
$\,\int_0^t \xi(t^\prime)\,dt^\prime\,$, %
in respect to real time. %

\,\,\,\, Similar separation of high-frequency noise and %
1/f-type noise appears also, in the course of %
naturalal approximations, in the microscopic %
statistical-mechanical theory of molecular %
Brownian motion (see e.g. \cite{p12,p0802,p1209,bkn}). %

\,\,\,\, At the same time, importantly, formula (\ref{f3_20}) %
can be derived without such separation, by using %
the equilibrium scale invariance reasonings plus  %
some FDR only (see \cite{bk3}).

\,\,\,
\end{thebibliography}
\end{document}